\documentclass[preprint]{emulateapj}

\shorttitle{CO in Sextans A}
\shortauthors{Shi et al.}

\begin{document}


\title{The Weak Carbon Monoxide Emission In An Extremely Metal Poor Galaxy, Sextans A \footnote{Based on observations carried out with the IRAM 30m Telescope.
IRAM is supported by INSU/CNRS (France), MPG (Germany) and IGN (Spain).}}


\author{Yong Shi\altaffilmark{1,2, 3}, Junzhi Wang\altaffilmark{4,5}, Zhi-Yu Zhang\altaffilmark{6}, Yu Gao\altaffilmark{7,5,3}, Lee Armus\altaffilmark{8}, George Helou\altaffilmark{8}, Qiusheng Gu\altaffilmark{1,2, 3}, Sabrina Stierwalt\altaffilmark{9}}

 \altaffiltext{1}{School of Astronomy and Space Science, Nanjing University, Nanjing 210093, China.}
 \altaffiltext{2}{Key Laboratory of Modern Astronomy and Astrophysics (Nanjing University), Ministry of Education, Nanjing 210093, China.}
 \altaffiltext{3}{Collaborative Innovation Center of Modern Astronomy and Space Exploration, Nanjing 210093, China.}
 \altaffiltext{4}{Shanghai Astronomical Observatory, Chinese Academy of Sciences, 80 Nandan Road, Shanghai 200030, China.}
 \altaffiltext{5}{Key Laboratory for Radio Astronomy, Chinese Academy of Sciences.}
 \altaffiltext{6}{Institute for Astronomy, University of Edinburgh, Royal Observatory, Blackford Hill, Edinburgh EH9 3HJ, UK}
 \altaffiltext{7}{Purple Mountain Observatory,  Chinese Academy of Sciences, 2 West Beijing Road, Nanjing 210008, China.}
 \altaffiltext{8}{Infrared  Processing and Analysis   Center,   California    Institute   of   Technology,   1200 E. California Blvd, Pasadena, CA 91125, USA.}
 \altaffiltext{9}{Department of Astronomy, University of Virginia, P.O. Box 400325, Charlottesville, VA 22904, USA.}

\email{yshipku@gmail.com}

\begin{abstract}

Carbon monoxide  (CO) is  one of  the primary coolants  of gas  and an
easily accessible tracer of molecular gas in spiral galaxies but it is
unclear if  CO plays a similar  role in metal poor  dwarfs. We carried
out a  deep observation with  IRAM 30 m  to search for CO  emission by
targeting the brightest  far-IR peak in a nearby  extremely metal poor
galaxy, Sextans A,  with 7\% Solar metallicity.  A  weak
CO J=1-0  emission is seen, which  is already faint enough  to place a
strong constraint  on the  conversion factor ($\alpha_{\rm  CO}$) from
the CO luminosity  to the molecular gas mass that  is derived from the
spatially resolved dust  mass map.  The $\alpha_{\rm CO}$  is at least
seven  hundred times  the  Milky  Way value.  This  indicates that  CO
emission  is  exceedingly  weak  in  extremely  metal  poor  galaxies,
challenging its role as a coolant in these galaxies.

\end{abstract}


\section{Introduction}

Stars  form out  of molecular  clouds \citep{Kennicutt98, Gao04}.  The
efficient cooling  of  molecular gas is  the prerequisite  for gas
collapse and star formation.  Among molecular species, carbon monoxide
(CO)  plays an  important role  in cooling  molecular gas  in  the low
temperature and density  regime \citep{Goldsmith01}.  The CO emission is
thus related to  the fundamental question about how  stars form out of
gas.  CO  cools gas by radiation,  and its bright  emission renders CO
the most common tracer  of the molecular gas mass \citep{Young82}, while
the two most  abundant components, H$_{2}$ and He,  cannot be observed
directly in  emission at the characteristic temperature of  molecular
gas clouds.  Over  the past decade, the CO  emission has been detected
in   galaxies  with   increasingly   lower  metallicity \citep{Israel97,
Taylor98, Leroy07, Leroy11}, down to 15\% Solar \citep{Elmegreen13}. The
conversion  factor  from  CO   to H$_{2}$
(referred as $\alpha_{\rm CO}$) has been constrained by comparing with
the molecular  gas mass inferred from other  methods, showing values
10-100 times  larger than the  Milky Way value \citep{Bolatto13}  in low
metallicity environments.  Probing  CO emission at lower metallicities
establishes  if  CO emission  can  be  an  efficient gas  coolant  and
effective tracer of  molecular gas in metal poor  galaxies both locally and
 in early Universe.

Sextans A is a dwarf irregular at 1.4 Mpc with oxygen abundance of 7\%
Solar  \citep{Pettini04,  Kniazev05}.   Its  proximity  increases  the
detectability of potential CO emission in the galaxy. It is one of a few
extremely  metal poor galaxies  whose molecular  gas masses  have been
estimated through the spatially resolved dust map \citep{Shi14}, arguably
the most accessible  way for cold gas mass  measurements in metal poor
galaxies    \citep{Bolatto13}.    The    data   are    presented    in
\S~\ref{observations}. The result and discussion are in \S~\ref{results}
and    \S~\ref{discussions},   respectively.    Conclusions    are   in
\S~\ref{conclusions}.

\section{Observations}\label{observations}

\begin{figure}
\epsscale{0.8}
\plotone{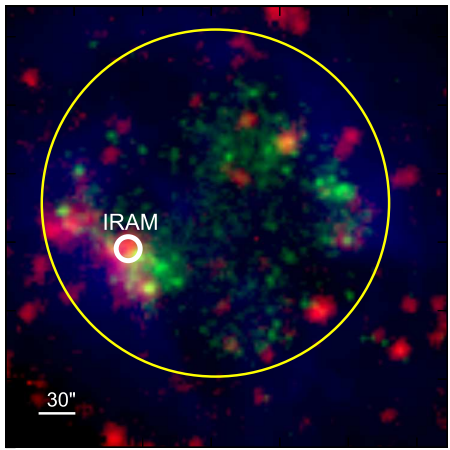}
\caption{\label{img}  False-color, multi-wavelength images  of Sextans
A.  Red is the sum of {\it Herschel} 160 and 250 $\mu$m data, green is
GALEX far-UV, and blue is radio  21 cm data showing the HI atomic gas.
The IRAM 30  m pointing at CO $J$=1-0 with the  beam size is indicated
by  the   white  circle,  while   the  yellow  circle   indicates  the
star-forming  disk  defined at  a  surface  brightness  of 25.9  ABmag
arcsec$^{-2}$ in the GALEX far-UV band.}
\end{figure}

The  CO $J$=1-0 observations  towards the  star formation  region (RA:
10:11:06.55,  DEC: -04:42:04.70, J2000)  in Sextans  A were  done from
Aug.   23th to  Aug  25th 2014,  using  the IRAM  30 meter  millimeter
telescope at Pico Veleta, Spain. The target region, which is 22 arcsec as 
the IRAM beam size at this frequency, is the far-IR peak
as  shown in  Fig. \ref{img}. 
    The  Eight  Mixer  Receiver  (EMIR)  with
dual-polarization  and  the   Fourier  Transform  Spectrometers  (FTS)
backend were  used, which  gave the frequency  channel spacing  of 195
KHz. The standard  wobbler switching mode with a  $\pm120''$ offset at
0.5  Hz beam  throwing was  used  for the  observations. Pointing  and
focusing  were  checked  about   every  2  hours  with  nearby  strong
millimeter emitting  quasi-stellar objects.   We read out  the spectra
every 2 minutes, while  the typical system temperature ($T_{\rm sys}$)
was about 280  K at this band.  Data reduction  was conducted with the
CLASS in GILDAS software package.  We checked each spectrum  and only used the
spectra  with $T_{\rm  sys}$ less  than  280 K  for final  discussion.
After throwing  out about 50\% spectra,  the total on  source time was
about 5 hours, which gave the noise level of about 4.1 mK in main beam
temperature  after  smoothing  the  frequency resolution  to  6.2  MHz
($\sim$ 16 km s$^{-1}$).

The {\it Herschel}  data at 70, 160, 250 and 350  $\mu$m, and the {\it
Spitzer}  data at  3.6, 4.5,  5.6,  8.0 and  24 $\mu$m  data are  from
\citet{Shi14} and \citet{Dale09}, respectively.  Because the dust
emission  within the  IRAM beam  is  partially resolved  instead of  a
simple point  spread function, to  measure the photometry  within the
IRAM beam, we first convolved  all images to the 350 $\mu$m resolution
based   on  the   convolution   Kernels  \citep{Aniano11}.    Aperture
photometry  at   all  bands  were   then  measured  at   this  spatial
resolution. The corresponding aperture correction factor is taken
as the ratio of the 70  um photometry at its native resolution to that
at the 350 $\mu$m resolution by assuming that the 70 $\mu$m map at its
native  resolution  can  resolve  the  dust  structure.  We  further
included the  far-ultraviolet (UV) images  from the {\it  GALEX} Space
Telescope archive  as well as maps  of atomic gas
\citep{Ott12}.

\begin{deluxetable*}{lcccccccccccccccccc}
\tabletypesize{\tiny}
\tablecaption{The Properties Of The Dwarf Galaxy Sextans A with the IRAM CO (1-0) Beam. }
\tablehead{
  \colhead{$I_{{\rm CO}(J=1-0)}$}   &  \colhead{$I_{{\rm CO}(J=2-1)}$}    &  \colhead{$L_{{\rm CO}(J=1-0)}$}          & \colhead{$M_{\rm dust}$}       & \colhead{$M_{\rm {H_{2}}+HI}^{\rm dust}$} &  \colhead{$M_{\rm {H_{2}}}^{\rm dust}$}   &  \colhead{$\alpha_{\rm CO(J=1-0)}$}              &  \colhead{SFR}                   &  \colhead{$L_{8-1000{\mu}m}$}   &  \colhead{$M_{\rm star}$} \\
   \colhead{mK km/s}              &  \colhead{mK km/s}               &  \colhead{K km/s pc$^{2}$}              & \colhead{M$_{\odot}$}         &  \colhead{10$^{7}$M$_{\odot}$}          &  \colhead{10$^{7}$M$_{\odot}$}          & \colhead{M$_{\odot}$/pc$^{2}$/(K km/s)}          &  \colhead{10$^{-4}$M$_{\odot}$/yr} &  \colhead{10$^{5}$L$_{\odot}$}  &  \colhead{$10^{5}$M$_{\odot}$}      
}
\startdata
145$\pm$39                        &  $<$ 460                         & 3670$\pm$990                           & 771$\pm$184                   & 1.1$\pm$0.3                          &   1.0$\pm$0.3                           & 2750$\pm$1110                                   &   2.2$\pm$0.4                     & 1.7$\pm$0.4                    & 1.7$\pm$0.4  \\   
\enddata
\end{deluxetable*}

\section{Results}\label{results}

\begin{figure}
\epsscale{0.8}
\plotone{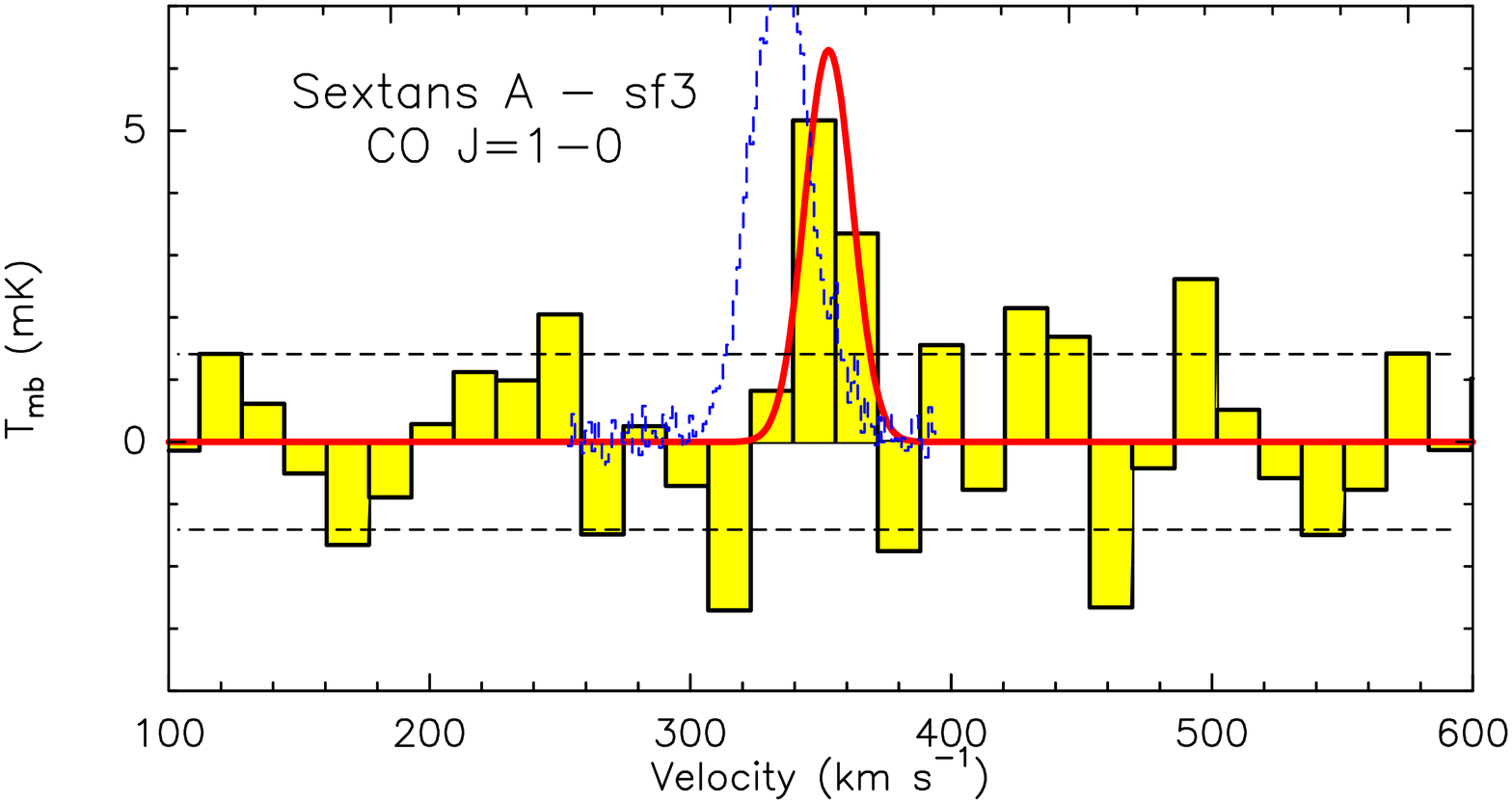} 
\caption{\label{CO_spectrum}  The submm spectrum  around the  CO J=1-0
frequency with source redshift $cz$=324 km/s. The red Gaussian profile
is  the fit to  the signal  at 350  km/s, which  gives signal-to-noise
ratio of 3.7 for the integrated  flux.  The two black dotted lines are
the 1-$\sigma$ noise levels. The  blue curve indicates the velocity of
the HI emission line within the IRAM beam as measured from the HI data
but with the absolute intensity scaled arbitrary \citep{Ott12}.  }
\end{figure}

\begin{figure}
\epsscale{0.8}
\plotone{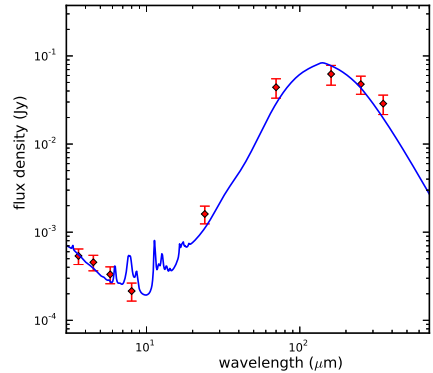} 
\caption{\label{dust_sed}  The Infrared  SED of  Sextans A  within the
IRAM  beam is  fitted to  derive the  dust mass.  Red symbols  are the
Spitzer and  Herschel photometric  points with 1-$\sigma$  error bars.
The  blue  solid  line  indicates  the  best-fit  by  the  dust  model
\citep{Draine07}.}
\end{figure}

\begin{figure}
\epsscale{0.8}
\plotone{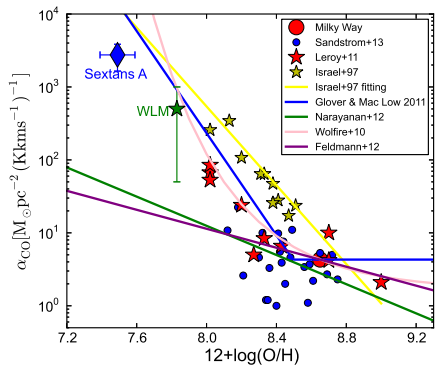} 
\caption{\label{alpha_CO}  The conversion  factor  ($\alpha_{\rm CO}$)
from the  CO luminosity  to the  molecular gas mass  for Sextans  A is
compared to  those at higher metallicities.   Symbols are observations
in the literature  \citep{Israel97, Leroy11, Sandstrom13, Elmegreen13}
where molecular  gas masses  are all based  on the  spatially resolved
dust maps except for WLM (see  text for the detail). Lines are models,
either    empirical    \citep{Israel97},    or   theoretical    models
\citep{Wolfire10, Glover11, Narayanan12, Feldmann12}. }
\end{figure}

Fig.  \ref{CO_spectrum}  shows the  observed  CO
$J$=1-0 spectrum  at a resolution of  16 km/s  with $cz$=  324 km/s.   A marginal signal  at 350  km/s was
seen.  A Gaussian fitting gives S/N  of 3.7 for the integrated flux, a
width  of  21$\pm$12 km/s  and  peak intensity  of  6.3  mK while  the
1-$\sigma$  noise is  1.4 mK.   The distribution  of the noise  level is
slightly  non-Gaussian. Using the negative values  of channels  over a
velocity range from  -500 to 1500 km/s, we  estimated a probability of
0.05\% for  the signal  to be artificial,  which corresponds to  a 3.4
$\sigma$  significance for  a  Gaussian distribution  of noises.   The
feature  lies within  the velocity  range of  atomic gas  as  shown in
Fig.  \ref{CO_spectrum}. The offset  in the  velocity between  the two
lines may be reasonable as the  two trace different phases of ISM.  As
listed in  Table 1, the intensity $I_{{\rm  CO}(J=1-0)}$ is 145$\pm$39
mK  km s$^{-1}$  in  T$_{\rm  mb}$, corresponding  to  a point  source
luminosity \citep{Gao04} of  $\sim$ (3.7$\pm$1.0)$\times$10$^{3}$ K km
s$^{-1}$  pc$^{2}$ for a  beam size  of 22  arcsec, much  smaller than
the previous  upperlimits \citep{Taylor98} of  a few$\times$10$^{4}$  K km
s$^{-1}$   pc$^{2}$   observed   at  the   position   (RA=10h11m07.3s,
dec=-4d42m22.5s,  J2000) near  our pointing  with  a beam  size of  55
arcsec.   The  3-$\sigma$  CO   $J$=2-1  assuming  the  same  velocity
dispersion as CO $J$=1-0 is 460 mK km/s.

The  conversion  factor  $\alpha_{\rm  CO}$  from  CO  to  H$_{2}$  is
quantified by  comparing the CO  luminosity to the molecular  gas mass
that is inferred from the spatially resolved dust map \citep{Israel97,
Leroy11, Sandstrom13}.  As  shown in Fig.~\ref{dust_sed}, the infrared
SED is first fitted with  the model of \citet{Draine07} as detailed in
\citet{Shi14}, based  on which the total cold  gas (HI+H$_{2}$) within
the  IRAM beam  is  measured by  multiplying  the dust  mass with  the
gas-to-dust ratio derived  from the diffuse regions of  the galaxy. As
discussed in  \citet{Shi14}, the  derived total cold  gas mass  is not
sensitive to the absolute dust mass but to the ratio of dust masses of
star-forming and diffuse regions. As  a result, as long as diffuse and
star-forming regions  have similar dust grains,  different dust models
gives a similar  estimate of cold gas masses within  the IRAM beam. To
derive the gas mass of star-forming regions, we do assume star-forming
regions have the same gas-to-dust  ratio as the diffuse regions, which
is  true for  spiral  galaxies \citep[e.g.][]{Sandstrom13}.   However,
this  assumption may  break down  in  dwarfs as  their star  formation
history is stochastic in both space  and time so that both diffuse and
star-forming regions are affected by different physical processes that
impact  the  gas-to-dust  ratio.   For  example,  the  SN  shocks  may
continuously destroy the  dust in the diffuse regions  in the past yet
meanwhile  dust  grains  in  dense  clouds  are  protected  from  such
destruction but instead grow  further. Nevertheless, such a hypothesis
needs systematic investigations before any conclusion can be made.

 After  subtracting the  atomic gas  mass from  the derived  total gas
mass, the molecular gas mass is shown to be about 10$^{7}$ M$_{\odot}$
as listed  in Table 1.  The  derived $\alpha_{\rm CO}$  value is about
(2.8$\pm$1.1)$\times$10$^{3}$    M$_{\odot}$     pc$^{-2}$    (K    km
s$^{-1}$)$^{-1}$, $\sim$ 700 times larger than the Milky Way
value   (4.3    M$_{\odot}$   pc$^{-2}$   (K    km   s$^{-1}$)$^{-1}$)
\citep{Bolatto13}.    In    Fig.~\ref{alpha_CO},   we   compared   our
measurement  to  those  in  the literature  \citep{Israel97,  Leroy11,
Sandstrom13} whose molecular gas masses are derived in the same way as
ours except  for the galaxy Wolf-Lundmark-Melotte \citep[WLM,][]{Elmegreen13}.  The gas-to-dust
ratio used  to derive  the $\alpha_{\rm  CO}$ of WLM  is based  on the
relationship    between     $\alpha_{\rm    CO}$    and    metallicity
\citep{Remy-Ruyer14}, which shows a  large scatter ($>$ 10) around the
metallicity of WLM.  The metallicity  measurement of Sextans A was not
taken exactly at  the CO pointing.  But as  no metallicity gradient is
seen in Sextans  A \citep{Kniazev05}, we did not  expect large offsets
($>$ 0.1  dex) from the value  quoted here.  Our  $\alpha_{\rm CO}$ is
much larger than those at  higher metallicities.  The overall trend of
$\alpha_{\rm CO}$  with metallicity  is very steep,  with a  power law
index between 2.5  and 3.  This implies that CO  emission is very weak
in  extremely  metal  poor  galaxies.   While  our  works  offer  very
interesting  constraints  on  the  $\alpha_{\rm CO}$  at  extreme  low
metallicity, the result is solely based on one region of $\sim$ 20 
arcsec compared to the entire galaxy of five arcmin.   Only  about 10\%  of  the total  molecular  gas  of Sextans  A
\citep{Shi14} is within the  beam. Large scatters of $\alpha_{\rm CO}$
may be expected  with future detections of CO  emission over different
regions of the galaxy.

The relative  strength of  the CO brightness  can also be  compared to
other quantities of Sextans A  measured within the IRAM beam as listed
in Table 1.  By integrating the  IR SED of the IRAM region, the 8-1000
$\mu$m infrared luminosity  of dust within the beam  is measured to be
1.7$\times$10$^{5}$ $L_{\odot}$. So the CO luminosity in K km s$^{-1}$
pc$^{2}$ relative  to the infrared luminosity in  $L_{\odot}$ is about
1:45, which is slightly  larger than spiral galaxies \citep{Genzel10}.
This is partly  caused by the rarity of both dust  and CO in extremely
metal poor  galaxies while different excitation mechanisms  for CO and
dust must  also play some roles.   If comparing to  the star formation
rate    (SFR)    as    measured    from    the    far-UV    luminosity
\citep{Kennicutt98b},  the SFR per  CO luminosity  is about  ten times
higher     than    star-forming     non-merger     massive    galaxies
\citep{Genzel10}.  If scaling  by the  stellar mass  as  measured from
Spitzer  3.6  and  4.5  $\mu$m measurements  \citep{Eskew12},  the  CO
luminosity per  stellar mass of Sextans  A within the IRAM  beam is on
average   10   times    lower   than   non-merger   massive   galaxies
\citep{Genzel10}.

\section{Discussions}\label{discussions}

The detection  of CO  $J$ 1-0 emission  in Sextans  A, if it  is real,
offers a crucial evidence for the existence of molecular gas in such a
low metallicity  environment, confirming the result  inferred from the
dust  measurement  \citep{Shi14}.   Although  warm  H$_{2}$  has  been
revealed by  infrared spectroscopic observations  in several extremely
metal poor  galaxies \citep{Hunt10}, CO emission,  unlike warm H$_{2}$
 that  traces shocked regions, is  known as tracers  of ISM where
stars form  in massive  metal rich galaxies.   The detection of  CO in
Sextans A  is a first step  toward determining whether CO  can serve a
similar role in extremely metal poor galaxies.

CO is one of few efficient  coolants of molecular gas in galaxies.  In
extremely metal  poor galaxies,  due to the  photo-dissociation,
the CO can only survive in  a tiny core of molecular gas clouds, while
H$_{2}$ can be self-shielded  and exist in thick envelopes surrounding
the CO  molecular core.   Thus cooling through  CO in  extremely metal
poor galaxies  may be not as  effective as in metal  rich galaxies. If
the  CO  molecular  cores  in  metal poor  galaxies  follow  the  same
relationship  between the viral  masses and  CO luminosities  of giant
molecular clouds  in Milky  Way and nearby  galaxies \citep{Solomon87,
Bolatto13}, the viral mass of the CO molecule core for the observed CO
luminosity  is estimated  to be  about  3$\times$10$^{4}$ M$_{\odot}$.
This is only  a tiny fraction of dust-based  total molecular gas mass,
as small as  0.3\%.  Then the measured  CO within  the IRAM  beam is  associated with  only  a small
fraction  of the  total molecular  gas within  the same  beam,  and it
cannot effectively  cool the bulk of  the molecular gas  in Sextans A.
 As a result of the photo-dissociation of CO, atomic  or  ionized 
carbon may  become  abundant.
Studies did show that [CII] 158 $\mu$m is much brighter relative to CO
in  dwarfs  as  compared  to in  spirals  \citep{Israel11,  Brauher08,
Madden13}.  With spatially resolved studies of IC 10, \citet{Madden97}
showed that the H$_{2}$  column density associated with ionized carbon
can be  five times the observed  HI density in order  to interpret the
cooling as implied by the [CII] luminosity, and the associated H$_{2}$
mass may be 100 times the mass of H$_{2}$ associated with CO. However,
unlike CO,  [CII] with  its high excitation  temperature (97  K) still
could  not cool  the  gas to  the  low temperature  at  which gas  may
contract to the high density for stars to form.

 Fig. \ref{alpha_CO}  further  shows the  comparison between  the
measured  extreme large $\alpha_{\rm  CO}$ of  Sextans A  with models'
predictions.    The   empirical   relationship   \citep[yellow   solid
line,][]{Israel97}  based  on galaxies  above  20\%$Z_{\odot}$ 
predicts that  $\alpha_{\rm  CO}$ $\propto$ $Z^{2.7}$,  giving a
value several  times larger than  our observed one  at 7\%$Z_{\odot}$.
Based on the principle that the CO abundance is primarily regulated by
photo-dissociation  and  the  H$_{2}$  is  self-shielded,  theoretical
models   of  \citet{Glover11}   and  \citet{Wolfire10}   predict  that
$\alpha_{\rm CO}$  is a strong  function of the visual  extinction for
individual clouds.  If assuming  the visual extinction is proportional
to the metallicity, $\alpha_{\rm  CO}$ then increases rapidly with the
decreasing metallicity  in both models, predicting  values larger than
the  observed for  Sextans A  as shown  in Fig.   \ref{alpha_CO}.  The
model of  \citet{Narayanan12} also employed  photo-dissociation region
models but  for the integrated galaxies instead  of individual clouds,
and found on  average a flat trend of $\alpha_{\rm  CO}$ as a function
of metallicity, producing  a value much smaller than  the observed one
at   7\%$Z_{\odot}$.   Unlike   the   above  models,   the  model   of
\citet{Feldmann12}    employs    small   scale    magneto-hydrodynamic
simulations combined with large-scale simulations of gas distributions
to  predict $\alpha_{\rm  CO}$ as  a function  of  metallicities.  For
physical scales  of our IRAM beam  size (sub-kpc), a  shallow trend as
indicated  by the purple  line in  Fig.  \ref{alpha_CO}  is predicted,
with predicted values significantly smaller than the observation.

As a  summary, current  models predict a  large range  of $\alpha_{\rm
CO}$ at the  metallicity of Sextans A, some  significantly larger than
the observed one  while some much smaller than  the observation. It is
still difficult to judge which model  is better or worse just based on
one data point, which is  also because many assumptions are invoked in
theoretical models in order to  produce the trend of $\alpha_{\rm CO}$
as a  function of metallicity.   A large range of  model's predictions
reflects the limited knowledge  about CO formation and destruction, as
well as properties of gas  clouds over different spatial scales at the
extreme low metallicity.  If $\alpha_{\rm  CO}$ is truly large as seen
in  this study,  CO may  be  not an  efficient coolant  in metal  poor
galaxies, and  its application as a  tracer of the  molecular gas mass
may be inappropriate in the early Universe.

\section{Conclusions}\label{conclusions}

We reported a  marginal detection of CO $J$=1-0  emission by targeting
the brightest far-IR  peak in an extremely metal  poor galaxy, Sextans
A. The signal is about 10 km/s offset from the HI peak within the same
beam.   The $\alpha_{\rm  CO}$  is further  derived  by comparing  the
estimated  CO flux  to  the H$_{2}$  mass  as inferred  from the  dust
map. In spite of a marginal  signal, the observation is deep enough to
play  a strong  limit on  the $\alpha_{\rm  CO}$ that  is  about seven
hundred times the  Milky Way value. This suggests  that CO emission is
exceedingly  weak in  extremely metal  poor galaxies,  challenging its
role as  a gas coolant  in these galaxies. Current  theoretical models
produce  a large  range of  $\alpha_{\rm  CO}$ at  the metallicity  of
Sextans A.

\acknowledgments

We thank the  anonymous referee for constructive comments  that improve the
paper significantly.  Y.S.  acknowledges  support for this  work from
Natural Science Foundation of China under the grant of 11373021 and by
the   Strategic   Priority  Research   Program   ``The  Emergence   of
Cosmological Structures''  of the  Chinese Academy of  Sciences, Grant
No.  XDB09000000. Y.G. is supported under grants of 11390373, 11420101002 \& 
XDB09000000.

\end{document}